\documentclass[fleqn,usenatbib]{mnras}

\usepackage{newtxtext,newtxmath}

\usepackage[T1]{fontenc}

\DeclareRobustCommand{\VAN}[3]{#2}
\let\VANthebibliography\thebibliography
\def\thebibliography{\DeclareRobustCommand{\VAN}[3]{##3}\VANthebibliography}

\usepackage{graphicx}	
\usepackage{amsmath}	
\usepackage{color}
\usepackage{comment}
\usepackage{threeparttable}
\usepackage{subfig}

\title[The stress-pressure lag in MRI turbulence]{The stress-pressure lag in MRI turbulence and its implications for thermal instability in accretion discs}

\author[L.E. Held et al.]{
Loren E. Held,$^{1,2}$\thanks{E-mail: loren.held@aei.mpg.de (LEH)}
Henrik N. Latter$^{1}$
\\
$^{1}$Department of Applied Mathematics and Theoretical Physics, University of Cambridge, Centre for Mathematical Sciences,\\
 Wilberforce Road, Cambridge CB3 0WA, UK\\
 $^{2}$Max Planck Institute for Gravitational Physics (Albert Einstein Institute), Am M{\"u}hlenberg 1, Potsdam 14476, Germany
}

\date{Accepted XXX. Received YYY; in original form ZZZ}

\pubyear{2020}

\begin{document}
\label{firstpage}
\pagerange{\pageref{firstpage}--\pageref{lastpage}}
\maketitle

\begin{abstract}
The classical alpha-disc model assumes that the turbulent stress scales linearly with -- and responds instantaneously to -- the pressure. It is likely, however, that the stress possesses a non-negligible relaxation time and will lag behind the pressure on some timescale. To measure the size of this lag we carry out unstratified 3D magnetohydrodynamic shearing box simulations with zero-net-magnetic-flux using the finite-volume code \textsc{PLUTO}. We impose thermal oscillations of varying periods via a cooling term, which in turn drives oscillations in the turbulent stress. Our simulations reveal that the stress oscillations lag behind the pressure by $\sim 5$ orbits in cases where the oscillation period is several tens of orbits or more. We discuss the implication of our results for thermal and viscous overstability in discs around compact objects.

\end{abstract}




\section{Introduction}
\label{INTRO}

The alpha-disc  model has underpinned the study of accretion discs for several decades \citep{1973shakura}. It has permitted researchers to develop semi-analytic theories with which to interpret observations and also direct numerical simulations. Despite its successes, this essentially mean field model makes strong assumptions that problematise its application to phenomena on lengthscales of order the disc scale height or shorter, and on timescales shorter than the viscous time \citep{1994balbus,ogilvie2003}. Instabilities, in particular, are potentially misrepresented, with the most famous example being thermal instability in radiation-pressure dominated accretion flows: while alpha-disc models predict instability and subsequent nonlinear oscillations \citep{lightman1974,shakura1976,1991honma,2001szuszkiewicz}, X-ray observations generally fail to find variability on the time-scales expected \citep[e.g.][]{Gierlinski2004}.

Several solutions have been proffered to explain the thermal instability conundrum \citep[e.g. see discussion in][]{ross2017}, but in this paper we focus on the time lag between variations in pressure and variations in the turbulent stress, an effect that can weaken the instability \citep{lin2011,ciesielski2012}. The alpha disc  prescription assumes that the stress responds \textit{instantly} to the pressure, but in reality it will take time for variations in one field to be communicated to the other through the complicated tangle of turbulent motions and magnetic fields. This lag is connected (and probably equal to) the stress's relaxation time (e.g. \cite{crow1968}), and if this is similar to the timescales of interest it may be necessary to solve a separate evolution equation for the stress itself, rather than adopt the oversimple alpha prescription \citep{kato1993,ogilvie2003,pessah2006}.  

In this paper we measure the relaxation time of the turbulent stress in local simulations of the zero-net-flux magnetorotational instability (MRI), with the finite volume Godunov code PLUTO \citep{mignone2007}. We construct numerical experiments in which we artificially drive oscillatory variations in pressure on various timescales, and then see how quickly or slowly the turbulent MRI stress responds. The choice to drive such cycles is not motivated by any particular astrophysical phenomena, but because such cycles make the calculation of the intrinsic lag easier.
Our results show that driving on intermediate to long timescales (of order or greater than the thermal time), the turbulent stress lags behind the pressure by only about 5 orbits.  If these results carry over to more realistic disc environments (especially those incorporating radiation pressure) then the lag is too short to impact on thermal instability unless the turbulent alpha is $\gtrsim 0.1$ (i.e. the dynamical and thermal times are similar), which can be the case in certain sub-states of X-ray binaries in outburst \citep[e.g.,][]{done2007}. The lag then weakens thermal instability, and in combination with other processes (magnetic fields, the stress's inherent stochasticity, etc.) may potentially suppress it \citep{Begelman2007,Oda2009, ross2017}. Finally, though only tangentially applicable to viscous overstability \citep{kato1978,blumenthal1984}, our simulations suggest that our measured lags may be more than sufficient to complicate its onset in MRI-turbulent gaseous discs \citep{ogilvie2001}.

The structure of the paper is as follows. In Section \ref{Background} we review the theoretical context and discuss thermal instability and viscous overstability. In Section \ref{METHODS} we discuss our set-up, initial conditions, and diagnostics. Section \ref{StressPressure_Results} presents our results. First we confirm the stress-pressure relationship reported in \cite{ross2016} in simulations that employ either no explicit cooling (heating runs), or otherwise constant-time-cooling (cooling runs). Next we show simulations with driven thermal oscillations in the mean pressure and study the resultant stress response, and any time lags that appear. Finally we discuss the results and conclude in Section \ref{StressPressure_Discussion}.

\section{Background}
\label{Background}

This section sketches out the context and motivation for this work, and explains some theoretical expectations and applications. 

\subsection{Pressure and stress variations}

In local simulations of the MRI the turbulent stress and the pressure can engage in a two-way interaction, but with each reacting upon the other on slightly separated timescales. 
On shorter times, of order an orbit or so, the MRI's turbulent stress undergoes stochastic fluctuations of non-negligible amplitude, and in particular in strong bursts. These bursts can be communicated to the
pressure via heating, in which case one observes the stress acting on the pressure \citep{hirose2009,latter2012}. But it is important to recognise that the stress fluctuations are essentially random, and cumulatively don't impart any mean trend if the gas is in thermal equilibrium. A slight lag exists between bursts in stress and bursts in pressure, because of the finite time it takes for the excess energy to tumble down the turbulent cascade and to thermalise.

If, however, on longer times there is a mean change in the average pressure, caused by an external mechanism or by instability for example, then the MRI's turbulent stress follows the pressure. This has been demonstrated in local simulations (of sufficiently large radial size) describing the pure heating or excessive cooling of the turbulent gas \citep{ross2016} and of thermal instability \citep{jiang2013,ross2017}. The link between pressure and stress these simulations show is bound up in the MRI saturation mechanism, though its exact details are yet to be fully understood. At the very least, it appears that compressibility limits the size of the largest MRI turbulent eddies and hence the magnitude of the subsequent stresses, though there remains evidence that the smallest scales are also involved \citep{fromang2007,Fromangetal2007,ross2016,ryan2017,ross2018}. 

It has yet to be established numerically that there exists a time lag in the second slower interaction described, i.e. between the response of the stress to the pressure. One might speculate that such a time lag should always be of order or somewhat longer than the dynamical timescale \citep{ogilvie2003}. As one expects the lag to be set by the MRI saturation mechanism, at the very least it should be bounded below by the longest eddy turnover time $\sim \Omega^{-1}$ and bounded above by the thermal time $\sim (\alpha\Omega)^{-1}$, where $\Omega$ is the local orbital frequency and $\alpha$ is the alpha parameter.

\subsection{Thermal instability}

We now show how a time lag in the stress impacts on thermal instability in disks \citep{lightman1974,shakura1976}, using a simple but transparent 1D toy model, based on analyses by \cite{lin2011} and \cite{ciesielski2012}.  A more complete account might include two coupled (possibly stochastic) equations for the stress and the energy \citep[cf.][]{hirose2009}.

Consider the energy equation for a specified blob of gas in the disk: $de/dt =\Gamma-\Lambda $, where $e$ is the blob's thermal energy, $\Gamma$ is its heating rate (proportional to the turbulent viscous stress), and $\Lambda$ is the cooling rate. Suppose the viscous stress depends on pressure (and thus $e$) -- but on its value a constant time $\tau$ in the past (as discussed in Section 2.1). It follows that the heating rate will also depend on the energy a time $\tau$ ago, and so we write $\Gamma =\Gamma[e(t-\tau)]$. We also take $\Gamma$ to be a monotonically increasing function of $e$. It is assumed, however, that the cooling depends on the \emph{instantaneous} thermal energy, and thus $\Lambda=\Lambda(e)$.

In equilibrium $\Gamma=\Lambda$ and $e=e_0$, a constant. This state is perturbed by a small purely thermal disturbance $e'$. After linearising the energy equation and taking $e' \propto \exp(s t) $, where $s$ is a growth rate, we obtain the dispersion relation
\begin{equation}
    s=\Gamma' \exp(-\tau s)-\Lambda'
\label{EQUN_Growthrate}
\end{equation}
where $\Gamma'= (d\Gamma/de)_0$ and $\Lambda'= (d\Lambda/de)_0$.

If there is no lag ($\tau=0$) in Equation \ref{EQUN_Growthrate}, then we have the usual result: $s= \Gamma'-\Lambda'$, with instability occurring when $\Gamma' > \Lambda'$. If satisfied, a small increase in energy will be reinforced because the change in heating outstrips the change in cooling; similarly, a small decrease in energy will be exacerbated. We hence denote $s_0=\Gamma'-\Lambda'$.

\begin{figure}
\begin{center}
\includegraphics[scale=0.39]{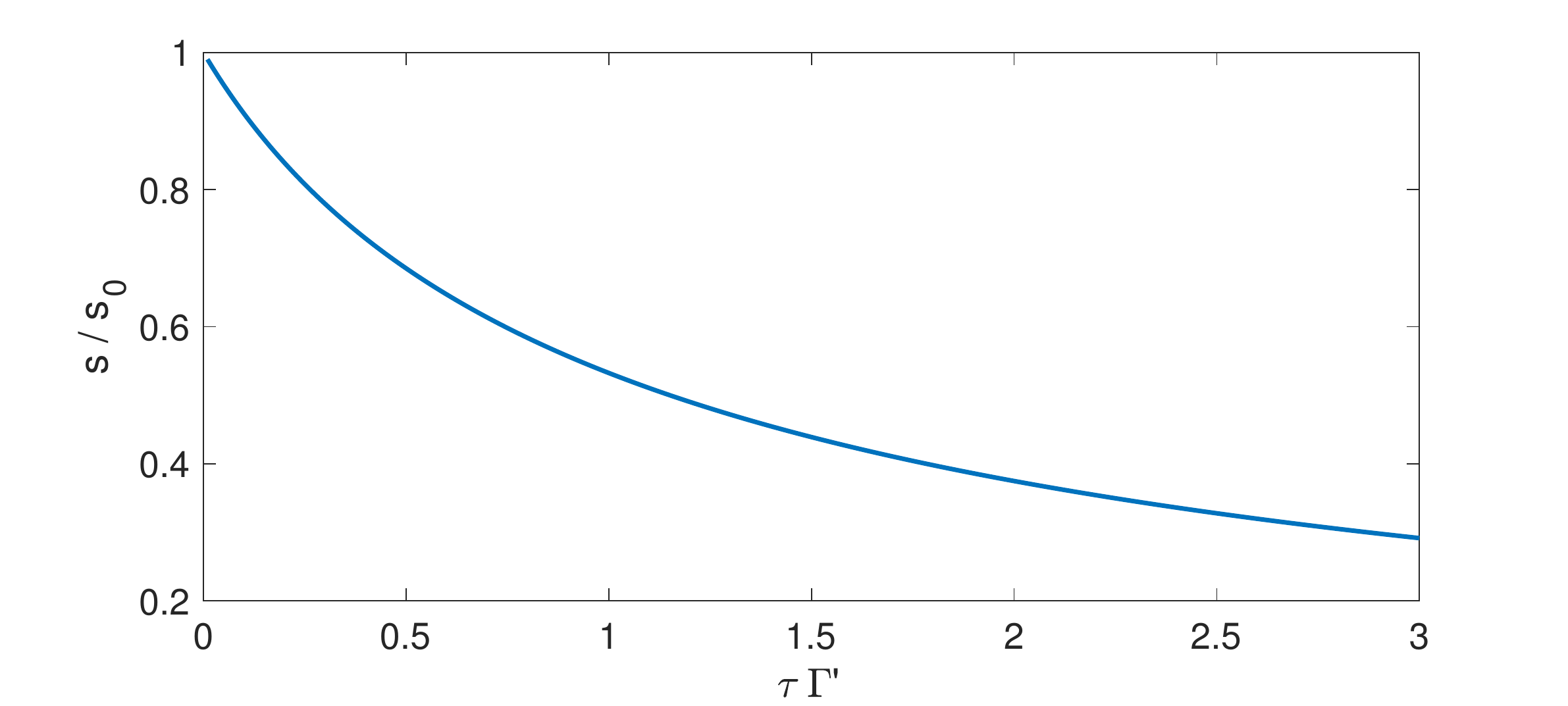}
\end{center}
\caption{Representative growth rate of thermal instability $s$ as a function of the time lag $\tau$. Here $s_0$ is the growth rate achieved with no lag, and $\Gamma'$ represents the inverse thermal timescale. We have chosen $\Lambda'/\Gamma'=1/2$ in \eqref{EQUN_GrowthrateLambert}. }
\label{growthrate}
\end{figure}

Next consider $\tau>0$. With a change of variable, 
the dispersion relation can be `solved':
\begin{equation}
s=  \frac{1}{\tau}\text{W}[\Gamma'\tau\text{e}^{\Lambda'\tau}] -\Lambda', \label{EQUN_GrowthrateLambert} \end{equation}
where W is the (first real branch of the) Lambert function \citep{corless1996lambertw}. To illustrate the effect of the lag on the growth rate of thermal instability, we consider two asymptotic limits of Equation \ref{EQUN_GrowthrateLambert}. First, we recognise that the thermal timescale is of order $1/\Gamma'$ and $1/\Lambda'$, and examine the limit of $\tau$ much less than the thermal timescale. After expanding $s$ in small $\Gamma'\tau$ and $\Lambda'\tau$, we obtain  
$s= s_0\left(1-\Gamma'\tau+\dots\right).$ 
As is clear, the time lag weakens the growth rate. If we next examine the opposite limit of a very long time lag $\Gamma'\tau\gg 1$, then we find
\begin{equation}
s \sim  \Gamma' \left(\frac{\log (\Gamma'\tau)}{\Gamma'\tau}\right) \ll s_0,
\end{equation}
and the growth rate is significantly reduced. Intermediate lags can also produce a non-trivial reduction in $s$, as can be observed in Figure \ref{growthrate}, which shows that $\tau$ of order the thermal timescale can halve the growth rate. The mathematics here is animating relatively straightforward physics: for example, a small increase in energy will induce a change in cooling, but the heating change (which formerly would outstrip the cooling) is delayed, and the thermal perturbation may return to equilibrium before it can be amplified.

Given the above analysis, we argue that the lag time $\tau$ poses a problem for thermal instability when it is of order or greater the thermal time $1/\Gamma'$, i.e. if $\Gamma'\tau \gtrsim 1$.
We next add to this the assumption $\tau \gtrsim \Omega^{-1}$ (cf. Section 2.1), i.e. the lag is naturally of order or greater than the dynamical time.  Combining the two estimates, yields a relatively straightforward (and rather strict) condition for the importance of the lag: the thermal time must be sufficiently close to the dynamical time. In a modified alpha-disc prescription with a retarded pressure dependence, we have $\Gamma'=\alpha\Omega$, and thus the condition for the lag's importance is roughly $\alpha>0.1$.

 Note that in previous local MRI simulations of thermal instability \cite[e.g.,][]{jiang2013,ross2017}) the measured alpha was $\sim 0.01$, and thus too small for the lag effect to be noticeable.\footnote{While there have been possible instances of thermal instability in global simulations, because only cooling runaways have been witnessed - and never a heating runaway - it is unclear if these correspond to true instability or simply absence of equilibrium \citep{mishra2016,skadowski2016}.} On the other hand, in order to match observed light curves of X-ray binaries we must have $\alpha\gtrsim 0.1$ in the outbursting state \citep{lasota2001}, which includes the radiation-dominated sub-states that should be thermally unstable \citep[e.g.][]{done2007}. It is in these contexts that we might expect the lag to reduce the instability growth rate and lengthen its characteristic timescale, though certainly not by more than an order of magnitude.     

\subsection{Viscous overstability}

Though this paper restricts itself to measuring the delay between pressure variations and stress variations, a lag in those quantities should be indicative of lags in strain and density as well, and more generally the stress's finite relaxation time. Thus a second application of our investigations is to the viscous overstability \citep{kato1978,blumenthal1984}, which is sensitive to a non-negligible stress relaxation. The viscous overstability is essentially a growing density wave \citep[f-mode;][]{ogilvie1998}: the wave produces an oscillatory perturbation in the stress, which extracts energy from the background orbital shear and directs it back into the wave. The instability can lead to eccentricity growth in both narrow rings and extended gaseous discs \citep{borderies1985,papaloizou1988,lyubarskij1994,ogilvie2001}, global oscillations in X-ray binaries \citep{chan2009,miranda2015}, and fine-scale structure in Saturn's rings \citep{schmit1995,Latter2009}. 

In order for there to be an appreciable energy flow from the background shear to the wave, the turbulent stress oscillation and the dynamical oscillation need to be sufficiently in phase: a significant lag between them can quench instability \citep{ogilvie2001,latter2006}. In dense planetary rings the two quantities respond almost instantaneously, as shown by N-body simulations and kinetic theory \citep{salo2001, Saloetal2001, latter2008}, and hence instability can proceed unproblematically. But, as discussed in Section 2.1, MRI turbulence will probably exhibit a relaxation/lag time of order or longer than the underlying inertial-acoustic oscillation, and this is likely to be sufficient to complicate the onset of overstability in turbulent gaseous discs \citep{ogilvie2001}.

\section{Methods}
\label{METHODS}

\subsection{Governing equations}
\label{METHODS_GoverningEquations}
We work in the shearing box approximation
\citep{goldreich1965,hawley1995,latter2017local},
which treats
a local region of a disc as a Cartesian box located at some fiducial
radius $r = r_0$ and orbiting with the angular frequency of the disc
at that radius $\Omega_0 \equiv \Omega(r_0)$. A point in the box has
Cartesian coordinates $(x, y, z)$ which point in the radial, azimuthal, and vertical directions.
The equations of magnetohydrodynamics in the box are
\begin{align}
&\partial_t \rho + \nabla \cdot (\rho \mathbf{u}) = 0, \label{SB1} \\
&\partial_t \mathbf{u} + \mathbf{u}\cdot\nabla \mathbf{u} = -\frac{1}{\rho} \nabla P - 2\Omega_0 \mathbf{e}_z \times \mathbf{u} + \mathbf{g}_{\text{eff}} + \frac{1}{\mu_0 \rho}(\nabla\times\mathbf{B})\times\mathbf{B} \nonumber \\ 
&\,\,\,\,\,\,\,\,\,\,\,\,\,\,\,\,\,\,\,\,\,\,\,\,\,\,\,\,\,\,\,\,\,\,\,\,\,+ \frac{1}{\rho}\nabla \cdot \mathbf{T}, \label{SB2}\\
& \partial_t (\rho e) +  \mathbf{u}\cdot \nabla (\rho e) = -\gamma \rho e \nabla\cdot\mathbf{u} + \frac{\eta}{\mu_0}|\nabla\times\mathbf{B}|^2+\Lambda ,
\label{SB3} \\
& \partial_t \mathbf{B} = \nabla\times(\mathbf{u}\times\mathbf{B})+\eta\nabla^2\mathbf{B}, \label{SB4}
\end{align}
with the symbols taking their usual meanings. We close the system with the caloric equation of state for a perfect gas $P = e
(\gamma - 1) \rho$ where $e$ is the specific internal energy. The adiabatic index (ratio of specific heats) is denoted by $\gamma$ and is taken to be $5/3$. Cooling is represented by  $\Lambda$, which takes prescriptions described in Section \ref{StressPressure_CoolingPrescription}.
All our simulations are unstratified, and the effective gravitational potential is embodied in the tidal acceleration $\mathbf{g}_{\text{eff}} = 2q\Omega_0^2x\hat{\mathbf{x}}$, where $q$ is the dimensionless shear parameter $q \equiv \left.d\ln{\Omega}/d\ln{r}\right\vert_{r=r_0}$. For Keplerian discs $q=3/2$ a value we adopt throughout. We employ explicit diffusion coefficients in our simulations. The viscous stress tensor is given by $\mathbf{T} \equiv 2\rho \nu \mathbf{S}$, where $\nu$ is the kinematic viscosity, and $\mathbf{S} \equiv (1/2)[\nabla \mathbf{u} + (\nabla \mathbf{u})^\text{T}] - (1/3)(\nabla\cdot\mathbf{u})\mathbf{I}$ is the traceless shear tensor. The explicit magnetic diffusivity (or `resistivity') is denoted by $\eta$.

\subsubsection{Cooling Prescription}
\label{StressPressure_CoolingPrescription}

In order to compare with previous results, some simulations adopt a power law cooling prescription,
\begin{equation}
\Lambda = -\theta P^m,
\label{EQUN_Cooling2}
\end{equation}
where $\theta$ and $m$ are parameters that can be adjusted to obtain a desired stable fixed point \citep[see][]{ross2017}. Unless stated otherwise, $m=2$ in all our simulations that employ this cooling prescription.

Our main runs investigate the lag between MRI turbulent stresses and pressure by employing a $\Lambda$ that drives thermal oscillations on long timescales. We expect the stress to be bursty on short dynamical times, but over the longer driven cycles it should be possible to observe a meaningful lag in the stress's response to the driven pressure. We deploy a piecewise time-dependent cooling function of the form
\begin{equation}\label{maincooling}
\Lambda = \begin{cases}
	-P / \tau_H,\,\,\,\,\,\,\text{(`long' timescale)}\\
	-P / \tau_C,\,\,\,\,\,\,\text{(`short' timescale)}
		\end{cases}
\end{equation}
where $\tau_H$ and $\tau_C$ are two different cooling times (with $\tau_H > \tau_C$). When the `long' cooling timescale $\tau_H$ is activated, heating by viscous dissipation (due to the MRI turbulence) overwhelms the cooling and the mean pressure tends to rise (the `heating phase', hence the `$H$' in $\tau_H$). Once the volume-averaged pressure exceeds some maximum critical value $\langle P \rangle = \langle P\rangle_{+}$ the cooling is switched to the `short' cooling timescale $\tau_C$, where the angle-brackets denote a volume-average (see Section \ref{METHODS_Diagnostics}). Now the cooling rate exceeds the turbulent heating rate and the mean pressure drops (the `cooling phase', hence the `$C$' in $\tau_C$). Once the mean pressure falls below some minimum critical value $\langle P\rangle = \langle P\rangle_{-}$ the cooling is switched back to the `long' cooling timescale and the cycle is repeated.

Using this simple time-dependent cooling prescription we can control the volume-averaged pressure on timescales comparable to $\tau_H$ and $\tau_C$, and in fact force it to oscillate between a minimum and maximum value of $\langle P \rangle_{+}$ and $\langle P \rangle_{-}$, respectively. Thus cooling times are free parameters which we vary in order to control the period of the pressure oscillations: these can be significantly longer than the orbital period or pushed to a timescale near an orbit. In practice, different combinations of $\tau_C$ and $\tau_H$ yield similar thermal oscillations. The choices that appear later in the paper were arrived upon by trial and error. 
 
\subsection{Numerical set-up}
\label{METHODS_NumericalSetUp}

\subsubsection{Code}
\label{Methods_Codes}
We use the conservative, finite-volume code \textsc{PLUTO} \citep{mignone2007} with the HLLD Riemann solver, 2nd-order-in-space linear interpolation, and the 2nd-order-in-time Runge-Kutta algorithm. In order to enforce the condition that $\nabla\cdot\mathbf{B}=0$, we employ constrained transport (CT), and the UCT-Contact algorithm to calculate the EMF at cell edges. To allow for longer time-steps, we take advantage of the \textsc{FARGO} scheme \citep{mignone2012}. When explicit viscosity and resistivity are included, we further reduce the computational time via Super-Time-Stepping (STS) \citep{alexiades1996super}. Ghost zones are used to implement the boundary conditions.

 \textsc{PLUTO} solves the governing equations in conservative form, not the primitive form given by \eqref{SB1}-\eqref{SB4}. Moreover, it evolves the total energy equation rather than the thermal energy equation. Note that due to the conservative implementation of the equations, kinetic and magnetic energy is not lost to the grid but converted directly into thermal energy.  
 In \textsc{PLUTO} the cooling term $\Lambda$ is not implemented directly in the total energy equation. Instead, it is included on the right-hand-side of the thermal energy equation, which is then integrated (in time) analytically.

\subsubsection{Initial conditions}
\label{METHODS_InitialConditionsAndUnits}
All our simulations are initialized from an equilibrium exhibiting spatially uniform
density and pressure profiles, $\rho_0$ and $P_0$. The background velocity is $\mathbf{u} = -(3/2) \Omega_0 x \,
\mathbf{e}_y$. At initialization we perturb all the
velocity components with random noise exhibiting a flat power
spectrum. The perturbations $\delta \mathbf{u}$ have maximum
amplitude of about $0.1\,c_{s0}$, unless stated otherwise. Here $c_{s0}=\sqrt{\gamma P_0/\rho_0 }$ is the sound speed at initialization.

All simulations are initialized with a \textit{zero-net-flux} magnetic field configuration: $\mathbf{B}_0 = B_0\sin{(k_x x)}\mathbf{\hat{e}}_z$. We take the radial wavenumber to be $k_x = 4(2\pi/L_x)$, where $L_x$ is the radial size of the numerical domain (see Section 3.2.4). The field strength $B_0$ is set by the ratio of gas pressure to magnetic pressure at the mid-plane, $\beta_0 \equiv P / (B_0^2 /2)$. In our simulations we set $\beta_0 \equiv 1000$. 

\subsubsection{Units}
 
 All quantities in the following sections are given in terms of dimensionless code units. The unit of time is selected so that $\Omega_0 = 1$, where from now on we drop the subscript on the angular frequency. The length unit is chosen so that the initial mid-plane sound speed $c_{s0} = 1$, which in turn defines a reference scale-height $H_0\equiv c_{s0} / \Omega_0=1$. Note, however, that the sound speed (and the scale-height) is generally a function of both space and time. Finally the mass unit is set by the initial density so that $\rho_0 = 1$.
 
\subsubsection{Diffusivities}

We employ both an explicit magnetic diffusivity of $\eta = 2\times10^{-4}$ and an explicit viscosity of $\nu = 8\times10^{-4}$. These values correspond to a Reynolds number of $\text{Re} = c_{s0}H_0/\nu = 1250$, a magnetic Reynolds number of $\text{Rm} = c_{s0}H_0/\eta = 5000$, and a magnetic Prandtl number of $\text{Pm} = \nu/\eta = 4$. The magnetic Prandtl number was chosen to be sufficiently high so that the MRI does not switch off \citep{fromang2007}. These values are the same as those employed in the $256^3$ simulations of \cite{ross2016} with explicit diffusion coefficients.

\subsubsection{Box size, resolution, and boundary conditions}
\label{METHODS_BoxSizeAndResolution}

Our boxes take a size of $L_x\times L_y \times L_z $, with $L_x=L_y=L_z=4H_0$, with a resolution
of $256^3$ (i.e. $64$ cells per scale height). We use shear-periodic boundary conditions in the $x$-direction
\cite[see][]{hawley1995} and periodic boundary conditions in the
$y$- and $z$-directions. In principle, mass should be conserved in this set-up, however we discovered that periodic boundary conditions in \textsc{PLUTO} paired with the \textsc{FARGO} algorithm  does lead to some very small mass loss ($<0.3\%$ in 500 orbits).

\subsection{Diagnostics}
\label{METHODS_Diagnostics}

The volume-average of a quantity $X$ is denoted $\langle X \rangle$ and is defined as 
\begin{equation}
\langle X \rangle(t) \equiv \frac{1}{V} \int_V X(x, y, z, t) dV
\end{equation}
where $V$ is the volume of the box.

In accretion discs, the radial transport of angular momentum is
related to the $xy$-component of the total stress
\begin{equation}
\Pi_{xy} \equiv R_{xy} + M_{xy},
\label{totalstress}
\end{equation}
in which $R_{xy} \equiv \rho u_x \delta u_y$ is the Reynolds stress, where $\delta u_y \equiv u_y + q\Omega x$ is the fluctuating part of the azimuthal velocity, and $M_{xy} \equiv -B_x B_y$ is the magnetic stress.

To calculate the cross correlation between the volume-averaged stress and pressure we first smooth the stress (which is quite bursty on timescales $\sim 1$ orbit) using a lag-free rolling window average  \citep[see][]{pandasRolling}. We then measure the cross-correlation (Pearson r) between the smoothed stress and the pressure by shifting the pressure time-series and repeatedly measuring the correlation between pressure and stress \citep[see][]{cheung2019}.

\section{Results}
\label{StressPressure_Results}

We first reproduce previous results by \citep{ross2016}, to make sure that our MRI turbulent stresses follow the pressure in pure heating and pure cooling runs. The rest of the section describes thermally driven runs, using the prescription in Eq.~\eqref{maincooling}, in which the pressure (and hence the stress) exhibits long time oscillations. Analysing these time series, we can characterise the lag between stress and pressure.

\subsection{Pure heating and cooling runs}
\label{StressPressure_HeatingRuns}

We first carry out several runs with heating only (i.e. $\Lambda=0$). In these simulations the pressure increases monotonically (due to MRI turbulent dissipation). Our aim is to check that the stress also increases and to measure its power law dependence on pressure, i.e. $\Pi_{xy} \propto P^q$.

Our fiducial heating simulation adopts the parameters presented in the last section. We find that its volume-averaged pressure increases by a factor of around 7 over 40 orbits, reaching $\langle P \rangle \sim 4.8$ just before the box-dominated regime begins (i.e. when the sound speed is of order or greater than $L_x\Omega$). The stress exhibits more complicated behavior, it fluctuates over short timescales ($\lesssim 1$ orbit), nevertheless it too clearly increases after the MRI's initial breakdown into turbulence.  Once the simulation reaches the box-dominated regime, the stress plateaus and remains quasi-steady in time, though still undergoing fairly large fluctuations of around 0.01 in magnitude. In summary, there is a correlation between the stress and pressure, as in \cite{ross2016}.

\begin{figure}
\captionsetup[subfigure]{labelformat=empty}
\subfloat[]{\includegraphics[scale=0.22]{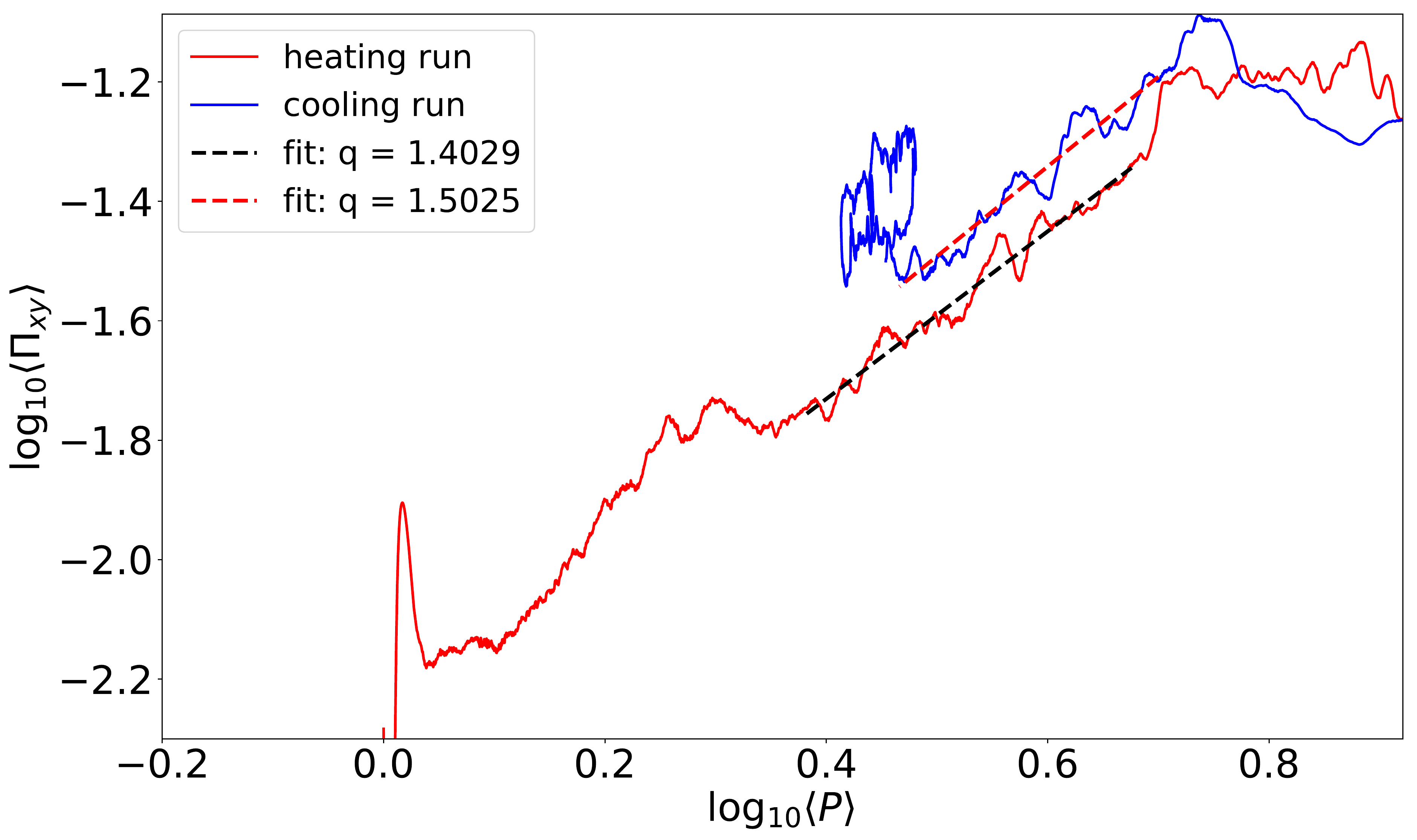}}
\caption{Log-log plot of total stress against pressure from a pure heating (red) and a pure cooling (blue) run. The slopes (dashed lines) are $q \sim$1.40 (heating) and $q \sim$1.50 (cooling).}
\label{FIGURE_StressPressureCoolingRunLogLog}
\end{figure}

To determine the stress-pressure scaling, we plot the stress and pressure against each other in 
Fig.~\ref{FIGURE_StressPressureCoolingRunLogLog} (the red curve). The curve, though complicated, can be fitted reasonably well with a power law of exponent $q\approx 1.4$, which is somewhat steeper than the slope of $q \approx 1$ measured in the non-ideal simulation of \cite{ross2016}. The exact relation is difficult to obtain, not least because it only appears in the transient phase of these heating simulations, and is thus subject to the initial condition and the simulation's numerical particulars. Moreover, the natural stochastic and bursty variation in the stress can partially destroy the correlation during the transient phase if one is unlucky.  To check on this we
carried out four additional heating runs with identical set-ups but different random initial conditions and calculated $q$s ranging from 0.7 to 1.8 (full details not shown). The average of the set is $\approx 1.1$.

Next we performed simulations that cooled monotonically in order to check that the stress follows the pressure not only when the latter increases but also when it decreases. All simulations were started from the box-dominated regime of our fiducial heating run (around orbit 50) and employed the non-linear cooling prescription given by Equation \eqref{EQUN_Cooling2}, with the cooling parameters set to $m=2$ and $\theta = 7.66\times10^{-3}$, respectively. An example cooling track is plotted in blue in Fig.~\ref{FIGURE_StressPressureCoolingRunLogLog}. For this example we find a power law exponent $q$ closer to 1.5, before the box settles on a stable fixed point in the thermal dynamics around $\langle P\rangle=2.51$.
This demonstrates that the stress depends on the pressure in a similar way both when the gas heats up and when it cools down. 

\subsection{Runs with oscillatory cooling}
\label{StressPressure_PeriodicCoolingRuns}

In this section we investigate how the stress responds to oscillatory variations in the mean pressure, and determine any time lags that appear. To do so we employ the explicit cooling prescription described in Section \ref{StressPressure_CoolingPrescription}. By adjusting the two cooling timescales $\tau_H$ and $\tau_C$ we can control the behavior of the mean pressure, allowing it to increase when the cooling timescale $\tau_H$ is sufficiently long, so that heating is greater than cooling (the `heating phase'), and to decrease when $\tau_C$ is sufficiently short, so that cooling overwhelms heating (the `cooling phase'). We present three different cases: a simulation in which the mean thermal (i.e. pressure) oscillation is `intermediate', i.e. exhibiting a period of $\sim 30-40$ orbits, and thus lying between the dynamical time ($\sim 1/\Omega)$ and thermal time ($\sim 1/(\alpha\Omega)\sim 100/\Omega$); `fast' oscillations, with period $\sim 10$ orbits, such that the stress cannot keep up with the changes in the pressure; and `slow' oscillations, with a period of order or greater than the thermal time.

\subsubsection{Intermediate thermal oscillations}

\begin{figure}
\captionsetup[subfigure]{labelformat=empty}
\subfloat[]{\includegraphics[scale=0.235]{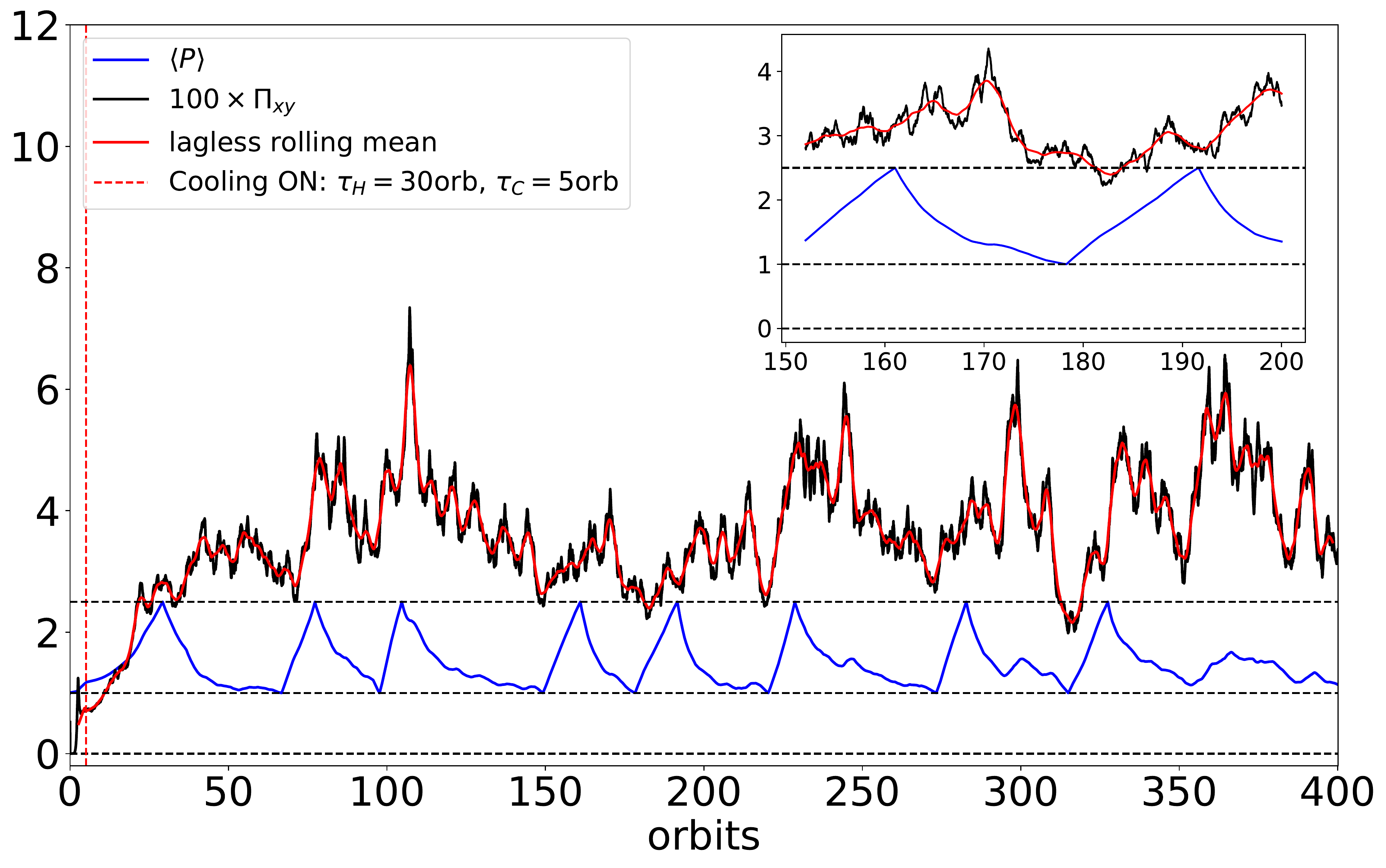}}
\caption{Time series of total stress and pressure from a  simulation exhibiting `intermediate' thermal oscillations. The black curve shows the volume-averaged stress (multiplied by $100$), the red curve shows the the lag-less rolling average of the stress, and the blue curve shows the volume-averaged pressure. The dashed vertical line shows the time at which cooling was turned on. The inset shows a time interval between orbit 150 and orbit 200: the lag can clearly be seen by eye.}
\label{FIGURE_PeriodicCoolingRes64NonIdeal_StressPressureTimeSeries}
\end{figure}

In Figure \ref{FIGURE_PeriodicCoolingRes64NonIdeal_StressPressureTimeSeries} we show the time-evolution of stress and pressure in our intermediate case. The cooling in this run was turned on shortly after the MRI's initial breakdown (orbit 5). The long and short cooling timescales were set to $\tau_H = 30$ orbits and $\tau_C = 5$ orbits, and the mean pressure was permitted to vary between $\langle P \rangle_{-} = 1$ and $\langle P \rangle_{+} = 2.5$. It is plotted in blue.
The black curve shows the stress normalized by $P_0$, and multiplied by 100 to make comparison with the pressure easier. To better track its longer timescale oscillations, we smooth out the shorter stochastic variation with a lag-free rolling mean (red curve).\footnote{For the rolling mean we employ a window size spanning 10k time steps, which is around 3.2 orbits.}

\begin{figure}
\captionsetup[subfigure]{labelformat=empty}
\subfloat[]{\includegraphics[scale=0.27]{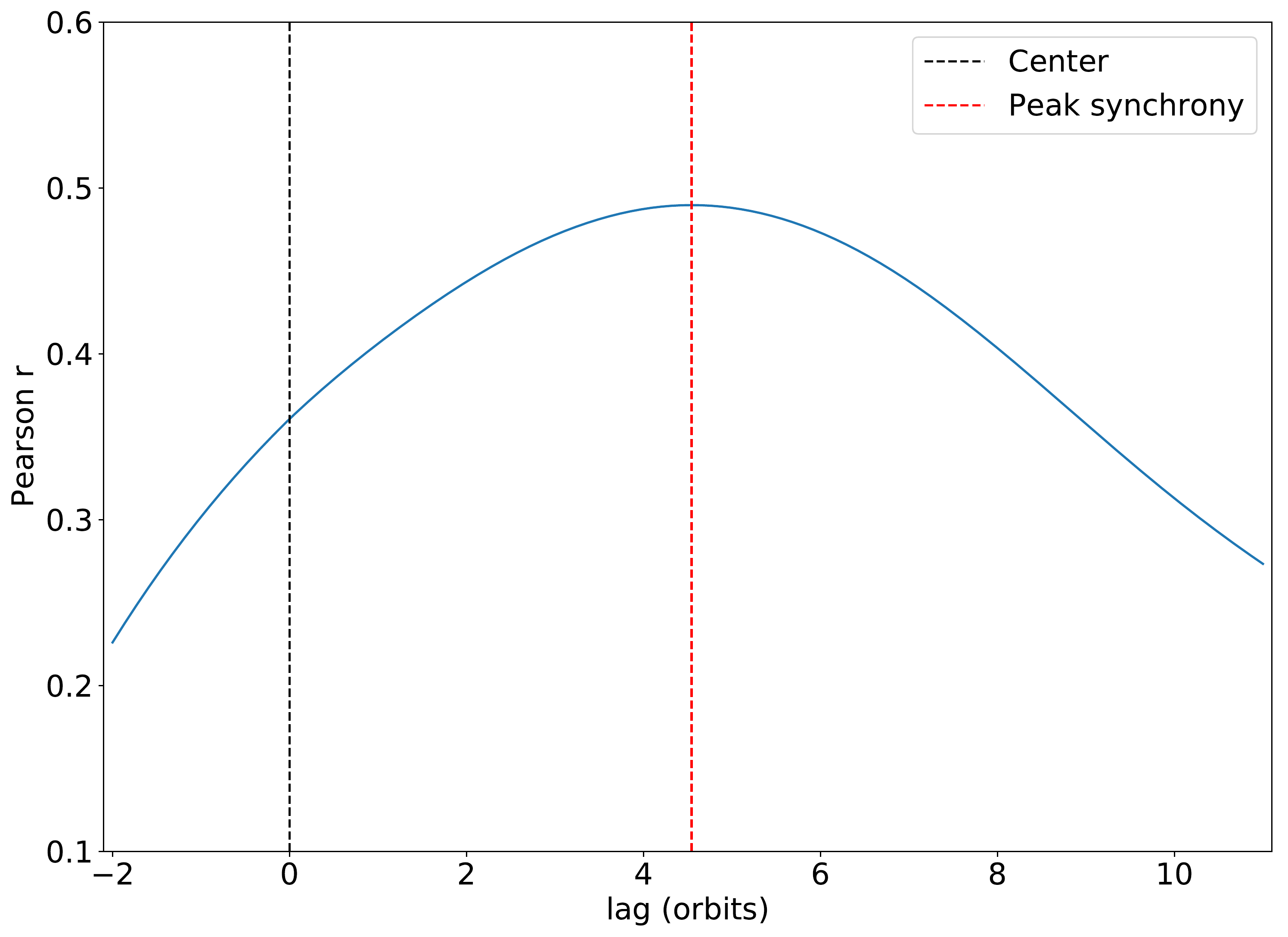}}
\caption{Cross-correlation coefficient between stress and pressure from the run exhibiting `intermediate' thermal oscillations. To the right of the black vertical dashed line is the correlation for a lag in which the stress follows the pressure, while to the left of the line the correlation is for the stress leading the pressure. The peak correlation (red vertical dashed line) occurs when the stress lags behind the pressure by about +4.54 orbits.}
\label{FIGURE_PeriodicCoolingRes64NonIdeal_CrossCorrelation}
\end{figure}

We begin by analyzing the time-series of the mean pressure. After the cooling function is turned on, the pressure continues to increase, though at a slower rate than before. At around orbit 25 the pressure reaches its maximum permitted value at which point the cooling timescale is switched to the shorter timescale $\tau_C$. The pressure consequently drops until $\langle P \rangle_-$, at which point the cycle repeats. On the other hand, after an initial transient (up to orbit 75), the stress tracks the pressure oscillation relatively closely, but with a noticeable lag of several orbits. This means during the start of each cooling phase the stress hits its peak, and because of the extra heating this provides, and the stress's inherent stochasticity, the cooling phases lengthen irregularly.

It is important to check whether our runs are in the box-dominated regime or not. In our heating-only run we found that the simulation entered the box-dominated regime once the stress reached around 0.06-0.07 (cf. the right-hand panel of Figure \ref{FIGURE_StressPressureCoolingRunLogLog}) after which the stress plateaued. In Figure \ref{FIGURE_PeriodicCoolingRes64NonIdeal_StressPressureTimeSeries} $\Pi_{xy}$ reaches $\sim 0.07$ only at the very tips of some of the outbursts. Thus we are confident that our oscillatory results are not contaminated by the box size.

To quantify the lag between the stress and the pressure we calculated the cross correlation between the two, shown in Figure \ref{FIGURE_PeriodicCoolingRes64NonIdeal_CrossCorrelation}. The correlation was calculated using data between orbits 66 and 350 in order to omit residual and transient behavior following initialization. We measure a peak correlation coefficient of $r = 0.47$. This peak occurs when the stress lags behind the pressure by around $4.54$ orbits.

\subsubsection{Fast thermal oscillations}

\begin{figure}
\captionsetup[subfigure]{labelformat=empty}
\subfloat[]{\includegraphics[scale=0.24]{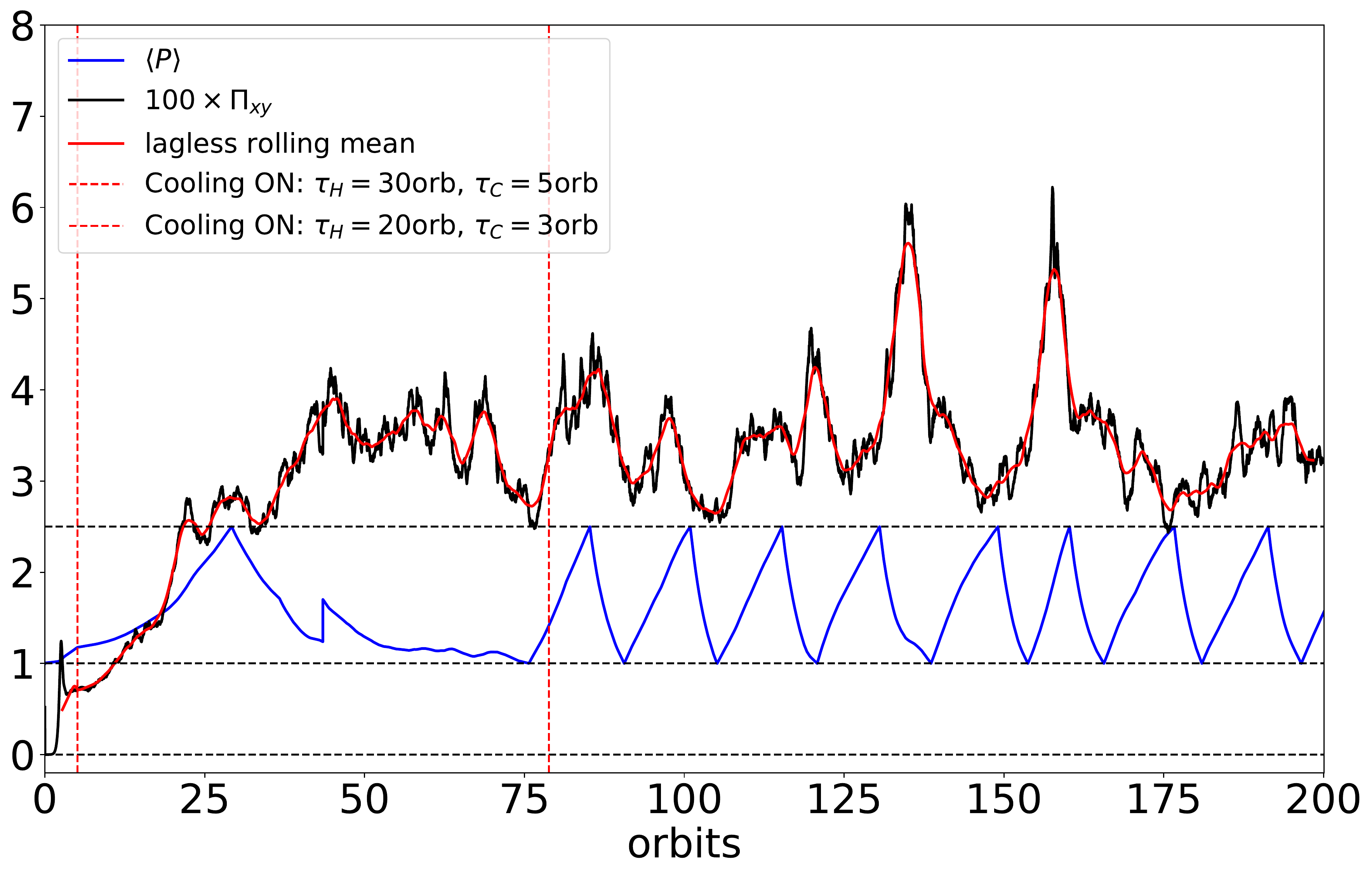}}
\caption{Time series of total stress and pressure from  simulation exhibiting `fast' thermal oscillations. The black curve shows the volume-averaged stress (multiplied by $100$), the red curve shows the lag-less rolling average of the stress, and the blue curve shows the volume-averaged pressure.  Note that after the second restart (vertical dashed red line) the stress is unable to keep up with the rapid variation in the mean pressure.}
\label{FIGURE_StressPressureCoolingTimeEvolution128Fast}
\end{figure}

Next we investigate how the stress tracks the pressure when the latter changes over timescales that are much shorter than those appearing in the `intermediate' case. To instigate such rapid variations we require that the pressure increases quickly during the heating-dominated phase (thus raising $\tau_H$), and that the pressure decreases quickly during the cooling-dominated phase (lowering $\tau_C$). Thus we set the cooling timescales to $\tau_H = 30$ orbits and $\tau_C=3$ orbits. This simulation is restarted from orbit 80 of our `intermediate' simulation, in order to avoid the initial transient phase, and was run until orbit 200. 

In Figure \ref{FIGURE_StressPressureCoolingTimeEvolution128Fast} we show the time-evolution of the total stress multiplied by 100, the rolling-averaged total stress, and the pressure. After the restart position (orbit 80) the mean pressure varies rapidly due to the shorter cooling timescale $\tau_C$. The stress struggles to track the pressure in this run, and after a few cycles the stress and pressure appear to be almost completely uncorrelated (e.g. from around orbit 120). This is confirmed by the cross-correlation (not shown), which shows that the Pearson r has no clear peak, and instead `wiggles' between small positive and negative values between $-15$ orbits to $+15$ orbits.

\subsubsection{Slow thermal oscillations}

\begin{figure}
\captionsetup[subfigure]{labelformat=empty}
\subfloat[]{\includegraphics[scale=0.24]{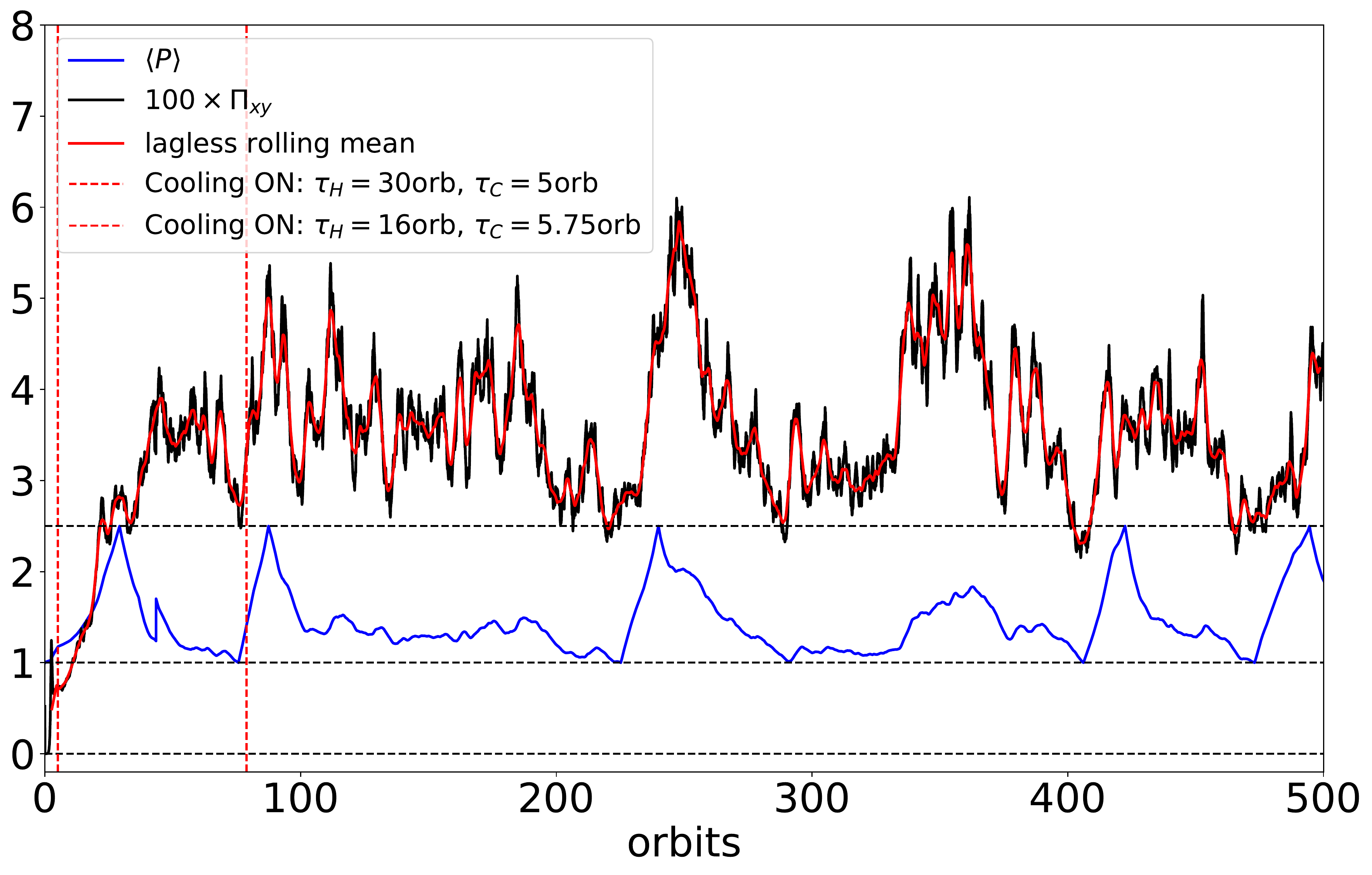}}
\caption{Time series of total stress and pressure from a simulation exhibiting `slow' thermal oscillations. The black curve shows the volume-averaged stress (multiplied by $100$), the red curve shows the lag-less rolling average of the stress, and the blue curve shows the volume-averaged pressure.}
\label{FIGURE_StressPressureCoolingTimeEvolution128Slow}
\end{figure}
\begin{figure}
\captionsetup[subfigure]{labelformat=empty}
\subfloat[]{\includegraphics[scale=0.27]{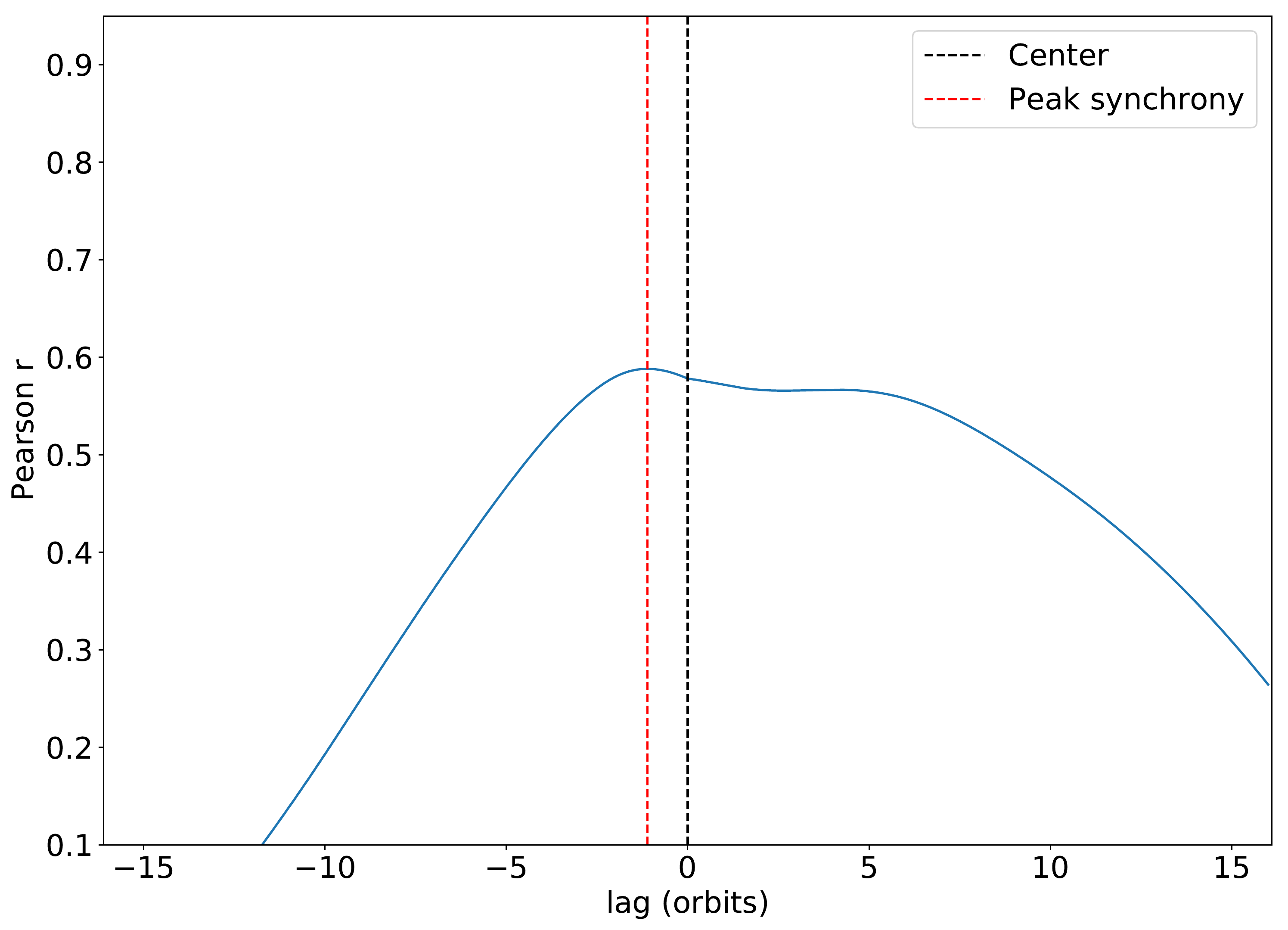}}
\caption{Same as Figure \ref{FIGURE_PeriodicCoolingRes64NonIdeal_CrossCorrelation} but the cross correlation is taken from a simulation with `slow' thermal oscillations.}
\label{FIGURE_StressPressureCorrelationSlow}
\end{figure}

Finally in our `slow' simulation we look at the other extreme: when the thermal driving occurs on a timescale of order the thermal time, and thus much longer than the dynamical timescale. This was achieved by lowering the `long' cooling timescale to $\tau_H = 16$ orbits (resulting in longer rise times of $\langle P \rangle$ during the heating-dominated phase) and increasing the `short' cooling timescale to $\tau_C = 5.75$ orbits (thus prolonging the decrease of $\langle P \rangle$ during the cooling-dominated phase). The simulation was run for 500 orbits.

 Our results appear in Fig.~\ref{FIGURE_StressPressureCoolingTimeEvolution128Slow}. The pressure undergoes an irregular, somewhat bursty, cycle of period $\sim 170$ orbits. The stress exhibits clear features of this thermal oscillation, and thus can be deemed to follow the pressure on this long timescale. As in previous simulations, the stress undergoes a great deal of large-amplitude stochastic variation on a range of shorter times. In fact, during the long cooling periods between heating bursts, we observe that the pressure experiences associated variations, because it is buffeted around by the stochastic heating. On these shorter times, the pressure possesses a random component that follows the stress, and in fact lags an orbit or so behind. 

To illustrate these two competing effects we plot the cross correlation of the two time series in Figure \ref{FIGURE_StressPressureCorrelationSlow}. We find significant correlation in a region encompassing a lag of zero, but the profile is rather blurred. We have a small peak at $-1.11$ orbits, which we associate with the component of the pressure that follows the stress's short-time stochastic variability during cooling phases. But this peak joins up to a plateau extending to a lag of about +6 orbits, a feature that we associate with the stress following the pressure on the long timescale of the driven thermal cycle. This `blurring' was not apparent in the intermediate case due to the dominance of the thermal oscillation, and it is likely that if we ran our `slow' cycle run for more periods we would observe a cleaner peak at a time lag of $\approx +5$.

\section{Conclusion}
\label{StressPressure_Discussion}

We have carried out unstratified shearing box zero-net-flux MRI simulations in \textsc{PLUTO}
in order to characterise the temporal relationship between turbulent stress and pressure, with an eye towards the existence, size, and sign of a time-lag in the variation of these two fields. There are several astrophysical applications of this work, but we have focused on its relevance for thermal instability and viscous overstability.

Our main set of simulations were run with a time-dependent cooling prescription which could drive long thermal oscillations in the gas, the period of which we could control. These would force oscillations in the pressure, which potentially could force oscillations in the turbulent stress, but with a lag we could measure. Simulations exhibiting thermal cycles on an intermediate timescale, with a period $\sim 50$ orbits, show the stress clearly lagging the pressure with a peak cross-correlation at about 4.5 orbits. Runs with a faster thermal cycle, periods $\sim 10$ orbits, possess a turbulent stress poorly correlated with the pressure: the stress struggles to `keep up' with the thermal oscillations because the measured time lag is too close to the oscillation period. Finally, when we attempted to drive long cycles the relationship between the stress and pressure is more complicated: while on the longer period of the cycle we find the stress following the pressure with a lag of $\sim 5$ orbits, as before, on shorter times the stress exhibits considerable stochastic variability, which the stress follows with a lag of around 1 orbit.  

The main application of these results is to thermal instability, particularly in low-mass X-ray binaries. Linear theory shows that its growth rate is significantly decreased if the time lag $\tau$ is of the order or greater than the thermal timescale, i.e.\ if $\tau\gtrsim (\alpha\Omega)^{-1}$ \citep[cf. Section 2.2;][]{lin2011,ciesielski2012}. If our numerical results generalise, and $\tau\approx 5$ orbits, then thermal instability is only impeded appreciably for large $\alpha \gtrsim 0.1$, which should indeed be the case in radiation-pressure dominated states in outburst \citep{lasota2001}. A precise quantitative estimate for the reduction in the thermal instability's growth rate is not possible to derive from our simulations. It is unlikely, however, that thermal instability is completely quenched by the lag; yet it could combine with other physics, such as the equilibrium's inherent stochasticity and magnetic fields \citep{Begelman2007,Oda2009,ross2017}, to delay the onset of thermal instability to the point that it is no longer dynamically relevant.
 An important caveat when applying our results is that our simulations were run with gas pressure only and, being zero-net flux, supported relatively low-level MRI turbulence, with $\alpha\sim 0.01$. Future work should attempt to generalise our calculations to conditions more representative of the inner radii of low-mass X-ray binaries in outburst.

Finally we note that the response of the stress to the pressure is only one aspect of the turbulent dynamics. More generally, the turbulent stress will depend on strain and density as well as pressure, and may relax towards a Navier-Stokes description on some characteristic timescale \cite[e.g.][]{ogilvie2001,ogilvie2003}. It is likely that the time-lags measured in our simulations correspond to this more general relaxation time. If such an identification can be made, then our results also bear on the onset of viscous overstability, and in fact indicate that the time-lags we find are sufficient to stabilise the disc.  

\section*{Acknowledgements}
This work was funded by a Science and Technologies Facilities Council
(STFC) studentship. Simulations were run on the Fornax cluster at the Department of Applied Mathematics and Theoretical Physics (DAMTP)
in Cambridge, and on the Sakura and Cobra clusters at the Max Planck Computing and Data Facility (MPCDF) in Garching. We would like to thank the referee for comments which helped us clarify certain concepts in the paper.

\section*{Data availability}
The data underlying this article will be shared on a reasonable request to the corresponding author.




\bibliographystyle{mnras}
\bibliography{2021StressPressurePaper_Bibliography} 



\bsp	
\label{lastpage}
\end{document}